\documentstyle[aasms4]{article}

\newcommand{\etal}{{\it et al.} }

\lefthead{Israel G.L. \etal}
\righthead{THE DISCOVERY OF 8.9S PULSATIONS FROM 2E0050.1--7247}
\received{03 -- Feb. -- 1997}
\accepted{13 -- May  -- 1997}
\slugcomment{\large Accepted for publication on Ap.J. Letters}
\begin{document}

\title{THE DISCOVERY OF 8.9 SECOND PULSATIONS FROM THE VARIABLE X--RAY SOURCE 
2E0050.1--7247 IN THE SMALL MAGELLANIC CLOUD }

\author{ {\bf G.L. Israel\altaffilmark{1,2,3}, L. 
Stella\altaffilmark{3,2}, L. 
Angelini\altaffilmark{4,5}, N. E. White\altaffilmark{4},  
P. Giommi\altaffilmark{6} and S. Covino\altaffilmark{7}  
}}

\altaffiltext{1}{International School for Advanced Studies (SISSA--ISAS), 
V. Beirut 2--4, I--34014, Trieste, Italy, gianluca@vega.sissa.it}

\altaffiltext{2}{Affiliated to the International Center for Relativistic 
Astrophysics}

\altaffiltext{3}{Osservatorio Astronomico di Roma, V. dell'Osservatorio 
2, I--00040 Monteporzio Catone (Roma), 
Italy,  stella@coma.mporzio.astro.it}
       
\altaffiltext{4}{Laboratory for High Energy Astrophysics, Code 662, NASA 
-- Goddard 
Space Flight Center, Greenbelt, MD 20771, USA,  
white@adhoc.gsfc.nasa.gov, angelini@lheavx.gsfc.nasa.gov}

\altaffiltext{5}{University Space Research Association}

\altaffiltext{6}{SAX Science Data Center, ASI, Viale Regina Margherita 
202, I--00198 Roma, Italy, giommi@sax.sdc.asi.it} 

\altaffiltext{7}{Osservatorio Astronomico di Brera, Via E. Bianchi 46, 
I--22055  Merate (Lecco), Italy, covino@merate.mi.astro.it}

\thispagestyle{empty}
\begin{abstract}

During a systematic search for periodic signals in a sample of ROSAT 
PSPC light curves, we discovered $\sim 8.9$~s X--ray
pulsations in 2E0050.1--7247,
a variable X--ray source in the Small Magellanic Cloud.  
The source was detected several times between 1979 and
1993 at luminosity levels ranging from $\sim$ 5 $\times$ 10$^{34}$ erg
s$^{-1}$ up to 1.4 $\times$ 10$^{36}$ erg s$^{-1}$ with both the
Einstein IPC and ROSAT PSPC.  The X--ray energy spectrum is consistent
with a power law spectrum which steepens as the source
luminosity decreases.  We revealed a 
pronounced H$\alpha$ activity from at least
two B stars in the X--ray error circles. 
These results strongly suggest that the X--ray pulsar
2E0050.1--7247 is in a Be--type massive binary.  
\end{abstract}
\keywords{binaries: general --- pulsars:  individual (2E0050.1--7247)
--- stars: emission--line, Be --- stars: rotation --- X--ray: stars}

\section{Introduction}

Accreting X--ray pulsars are mainly found in massive binaries
containing an OB donor star and X--ray transient activity is fairly
common among those systems which host a Be star.  The number of X--ray
pulsars of this class that are found in the Magellanic Clouds (MCs) is
small, but recent ROSAT observations have substantially increased the
sample (Hughes 1994; Dennerl \etal 1995; McGrath \etal 1994).  
Compared to those in our galaxy, massive
X--ray binaries in the MCs are usually characterised by a lower
galactic interstellar attenuation in their direction and a better
determined, though larger, distance. For three
X--ray pulsars in the MCs (SMC~X--1, A0538--66 and RX~J0059.2--7138)
the presence of a soft X--ray component, in addition to the
characteristic hard power law with exponential cutoff spectrum, has
been firmly established (Marshall, White \& Becker 1983; Mavromatakis
\& Haberl 1993; Hughes 1994; Corbet \etal 1995; Campana 1996). 

Based on several ROSAT PSPC observations, we report here the discovery
of 8.9~s X--ray pulsations from 1WGA J0051.8--7231, a
variable X--ray source in the SMC, which was tentatively identified 
(Bruhweiler \etal 1987) with the B star AV111.  The X--ray spectrum
showed a marked softening when the flux was a factor of $\sim 20$
below its peak value. We report also the detection of H$\alpha$
activity from AV111 and another B star within 
the X--ray error circles. Based on its position the latter star provides 
a more likely optical counterpart than AV111.

\section{ROSAT PSPC observations}

The Position Sensitive Proportional 
Counter (PSPC, 0.1--2.4~keV) in the focal plane of the 
X--ray telescope on board ROSAT observed the SMC field centered on 
RA = 00\arcdeg\ 54$^m$ 28$^s$.7 and DEC = --72\arcdeg\ 45$\arcmin$ 
36$\arcsec$.6 (equinox 2000)
on 1993 May 9--12 (seq. 600453, 17600~s exposure; Table\,1). 
The brightest source detected within the field was 1WGA J0051.8--7231 
at a position of RA = 00\arcdeg\ 51$^m$ 53$^s$.3, DEC = --72\arcdeg\ 
31$\arcmin$ 25$\arcsec$.3   
(uncertainty radius of $\sim 30\arcsec$). 

The ROSAT event list and spectrum of 1WGA J0051.8--7231 were
extracted from a circle of $\sim 1.0$$\arcmin$ radius 
(corresponding to an encircled energy, EE, of about 90\% ) around the X--ray 
position. Out of the 1750 photons contained in the circle we estimated 
that $\sim$ 100 photons derived from the local background around the source.  
The May 1993 light curve of 1WGA J0051.8--7231 was first analysed as 
part of a systematic study aimed at revealing periodicities in X--ray 
light curves of a sample of $\sim 23000$ X--ray sources selected from 
the White, Giommi \& Angelini catalog (1994; Israel 1996). 
The photon arrival times were corrected to the barycenter of 
the solar system and a 0.58~s 
binned light curve accumulated. The corresponding power spectrum calculated 
over the entire observation duration ($\sim 3.5$~d) is shown in Fig.~1a, 
together with the preliminary peak detection threshold 
described by Israel \& Stella (1996). Significant peaks were clearly  
detected around $\sim 2.5\times 10^{-3}$~Hz and $\sim 0.113$~Hz. 
While the former peaks 
are due to the wobble of the pointing direction ($\sim 402$~s 
period), the multi--peaked structure   
around a frequency of 0.1126~Hz is unique to 1WGA J0051.8--7231 (Fig.~1b); 
the highest of these peaks has a significance of $\sim 6.5 \sigma$ over the 
entire sample of ROSAT light curves analysed.
The power spectrum of the window function is shown in Fig.~1c with the same 
frequency scale as Fig.~1b; the peaks around the central 
frequency of $0.1126$~Hz are the sidelobes due to the 
satellite orbital occultations. 
In the absence of a unique phase fitting solution, we determined the best 
period in each of four consecutive time intervals by using a Rayleigh 
periodogram (cf. Leahy \etal 1983). The average of these periods gave
8.8816$\pm$0.0002 s (90\% uncertainties are used throughout this 
letter)\footnote{This value and error revise the preliminary 
ones reported by Israel \etal (1995).}.
An upper limit to the period derivative of $|$\.{P}$| <$ 3 $\times$ 
10$^{-10}$~s~s$^{-1}$ (3$\sigma$ confidence level) was derived.
The modulation was fairly sinusoidal, with a pulse fraction 
(semiamplitude of modulation devided by the mean source count rate) 
of $\sim 25$\% in the $0.1-2.4$~keV band (Fig.~1a). 
The arrival time of the pulse minima (that we adopted as phase 0) was 
determined to be JD 2449118.573193 $\pm$ 0.000003.

The PSPC energy spectrum was fitted with a simple power law model 
(Fig.~2a and Table~2). The best fit was obtained for a photon index of 
$\Gamma = 1.1\pm0.2$ and a column density of $N_H = (8\pm^3_2) \times$ 
10$^{20}~$~cm$^{-2}$ (the galactic hydrogen column in the direction of the 
SMC is $\sim 6 \times 10^{20}$~cm$^{-2}$). 
The 0.1--2.0~keV unabsorbed X--ray flux was 
$F_x\sim 2.5 \times 10^{-12}$~erg~cm$^{-2}$~s$^{-1}$, corresponding to 
a luminosity of $L_x\sim 1.2 \times 10^{36}$~erg~s$^{-1}$ for a distance 
of 65~kpc. 

The position of 1WGA J0051.8--7231 was included in 
several ROSAT PSPC fields observed between 1991 and 1993 
(seq. 600195, 600196, 500249n00 and 500251n00). 
The first observation was carried out on 1991 Oct 8--9 (1700~s exposure). 
The only detected source compatible with the position of 1WGA J0051.8--7231 
is 1WGA J0051.7--7231 (RA = 00\arcdeg\ 51$\arcmin$ 47$\arcsec$.7, DEC = 
--72\arcdeg\ 31$\arcmin$ 29$\arcsec$.2, uncertainty radius of 50$\arcsec$\,; 
Table\,1). 
The 1991 Oct 8--9 ROSAT event list and spectrum of 1WGA J0051.7--7231 
included about 370  photons accumulated from a circle of $\sim$ 
1.7$\arcmin$ radius (EE of  $\sim 90$\%) after background correction. 
The source count rate was estimated to be (6.9 $\pm$ 1.5) $\times$ 
10$^{-3}$~counts~s$^{-1}$ after correction for PSF and vignetting .  
The photon arrival times were corrected to the barycenter of the solar 
system and a search for coherent periodicities was than performed in 
a narrow range of trial periods (8.7--9.1~s) centered around the May 1993 
period. No significant peaks were found above the $3\sigma$ detection 
threshold. The corresponding upper limit on the pulsed fraction was 
$\sim 60$\%. A simple power law model produced 
an acceptable fit to the spectrum from the 1991 Oct 8--9 pointing 
(Fig.~2a and Table~2). The best fit parameters were determined to be 
$N_H = (7\pm^9_3) \times10^{20}~$cm$^{-2}$ 
and $\Gamma =2.9 \pm 0.8$. 
The corresponding 0.1--2.0~keV unabsorbed flux was about  
$F_x \sim 1 \times 10^{-13}$~ergs~cm$^{-2}$ s$^{-1}$, 
converting to a luminosity of $L_x\sim 5.0 \times 10^{34}$~erg~s$^{-1}$. 

In the remaining four ROSAT PSPC observations the source was not detected
and only upper limits on the count rates were determined (see Table\,1 for 
details).
In the ROSAT All Sky Survey (RASS) the region around the source was observed
on 1990 Oct 22--26 (667~s exposure). 
The source was not detected (2$\sigma$ upper limit of 
$1.1 \times 10^{-2}$~cts~s$^{-1}$\,; cf. Kahabka \& Pietsch 1996).

\section{Einstein IPC observations}

The SMC was extensively surveyed with the Einstein Observatory. The 
position of 1WGA J0051.8--7231 was included in three Imaging Proportional 
Counter (IPC, 0.15--4.5 keV) sequences (593, 6755 and 7988). 
These pointings were analysed in detail (Seward \& Mitchell 
1981; Bruhweiler \etal 1987; Wang \& Wu 1992). Two Einstein sources 
with compatible positions were detected within the ROSAT error circle of 
1WGA~J0051.8--7231: 2E0050.1--7247 (seq. 6755, source \#~27 of Wang \& 
Wu 1992) and 2E0050.1--7248 (seq. 7988, source \#~3 in Bruhweiler \etal 
1987). In the following we will use only the designation 2E0050.1--7247. 
The count rate of 2E0050.1--7247 increased by a factor of $\sim 5$ 
from 1980 Apr 15 (seq. 6755) to Apr 20 (seq. 7988, 
Table\,2; Wang \& Wu 1992). 
Correspondingly the hardness ratio underwent a significant increase.  
The source was not detected in sequence 593 
(upper limit of $\sim 2 \times 10^{-2}$~cts~s$^{-1}$; Wang \& Wu 1992). 

We reanalysed the Einstein IPC sequences 6755 and 7988 to search for the 
8.9~s periodicity. The IPC event lists were extracted from 
a circle of $\sim 1.6$$\arcmin$ radius (corresponding to an EE of $\sim 90$\%) 
around 2E0050.1--7247 and contained, respectively, $\sim 300$ and 100 
source photons. The arrival times of the 0.15--4.5~keV photons 
were corrected only for the earth motion; the search for coherent 
periodicities was then performed over the $\sim$ 8.65--9.15~s period range. 
No significant periodic signal was detected 
(upper limit of $\sim 75$\% and $> 100$\% for sequence 6755 and 7988, 
respectively).  
We fitted the IPC spectra with a power law model, keeping $N_H$ 
fixed at $8 \times 10^{20}$~cm$^{-2}$ (the best value 
obtained with the ROSAT PSPC in May 1993). The derived photon 
index is in both cases consistent with $\Gamma\sim 1$  
(large uncertainties are due to poor statistics; Table~2). 
The inferred 0.15--3.5~keV unabsorbed luminosities 
were $L_x\sim 1.4\times 10^{36}$ and $\sim 3.1 \times 10^{35}$~ergs~s$^{-1}$
for sequence 6755 and 7988, respectively. 

\section{Optical Observations}

The Einstein error circles of 2E0050.1--7247 include the optical
position of AV111 ($<$ 40$\arcsec$ offset), a variable 
B1 giant in the SMC, that was tentatively identified as the optical
counterpart (Bruhweiler \etal 1987; Wang \& Wu 1992). 
In order to reveal the H$\alpha$ active stars 
in the Einstein and ROSAT error circles ($\sim$40$\arcsec$ and 
$\sim$30$\arcsec$ radius, respectively), we obtained three $\sim 4$
$\arcmin$$ \times 4$$\arcmin$ images centered on AV111
with the ESO--La Silla Dutch 91\,cm telescope on 1996 Aug 12--13.
The exposure time was 90, 120 and 900\,s for
the V, B and H$\alpha$ frames, respectively. 
Comparing, for each object in the X--ray error circle, fluxes in the B,
V and $H_\alpha$ bands, we found several $H_\alpha$ bright stars. 
In particular, the two most prominent were AV111 itself and a previously 
unclassified star (hereafter Star1) at an optical position R.A. = 00\arcdeg\ 
51$\arcmin$ 56$\arcsec$.6, DEC = --72\arcdeg\ 31$\arcmin$ 30$\arcsec$.0 
(uncertainty $<$ 10$\arcsec$; see also Kahabka \& Pietsch 
1996). While consistent 
with the Einstein IPC error circles ($\sim$40$\arcsec$ radius), the position 
of AV111 was determined to lie $\sim$ 10$\arcsec$ outside the ROSAT PSPC error 
circle ($\sim$30$\arcsec$ radius; Fig.~3). On the contrary the position of 
Star1 is well within the PSPC error 
circle. The magnitude (V$\sim$13.4) and the color (B--V$\sim$ --0.28) were also 
compatible with those of a B giant at the distance of the SMC.

A medium resolution spectrum of AV111 was also taken 
on August 13, 1996  (600~s exposure) with the 
ESO--La Silla  1.5~m Spectroscopic Telescope. 
The spectral range and resolution were $\sim 4950-6950$\,\AA \, 
and $\sim 1$\,\AA \, (at 6500\,\AA). 
A pronounced H$\alpha$ emission feature was apparent in 
the spectrum with an equivalent width of $5.4 \pm 0.5$\,\AA\, 
We determined a line centroid to be at $6566$\,\AA\, corresponding to a 
radial velocity of $\sim 150$\,km\,s$^{-1}$ (the SMC line of sight velocity 
is $\sim 168$\,km\,s$^{-1}$s). 
Two distinct components with comparable intensities 
could be easily distinguished close to the line centroid.
These were peaked at $6564$ and $6569$\,\AA\, respectively, corresponding 
to a velocity difference of $\sim 200 \div 250$~km~s$^{-1}$.

\section{Discussion}

The 8.9~s pulsations from 2E0050.1--7247 together with the probable
identification of an H$\alpha$ emitting counterpart strongly suggest
that the X--ray source is an accreting magnetic neutron star likely in
a Be star X--ray binary. The X--ray source was detected over a range of
0.1--3.5~keV luminosities from $\sim 5 \times 10^{34}$ to $\sim$ 1.4
$\times 10^{36}$~ergs~s$^{-1}$ ;  on other occasions the source
remained undetected, with upper limits that in two cases were just
below the minimum detected luminosity.  These large variations are
suggestive of a transient behaviour. 

The relatively low peak X--ray luminosity detected from 2E0050.1--7247
indicates that the source was observed either during the decay (or the
rise) of a large outburst or during a lower luminosity (possibly
recurrent) outburst.  Increased H$\alpha$ emission from
AV111 or the other candidate (Star1) could provide
a useful indicator of resumed X--ray activity from 2E0050.1--7247
(this might be also useful to confirm the association of either candidates 
to the X--ray source). Based on the orbital period / spin period
correlation found by Corbet (1984) for Be star X--ray pulsars, the orbit
of 2E0050.1--7247 should be in the $\sim 30$~d period range. 

Despite the limited energy range, the power law spectral slope of
2E0050.1--7247 measured by ROSAT when the source luminosity was 
$\sim 10^{36}$~ergs~s$^{-1}$ ($\Gamma \sim  1$) is similar to that of most
accreting X--ray pulsars. A substantially softer spectrum ($\Gamma \sim3$) 
was measured when the source luminosity was $\sim 5\times
10^{34}$~ergs~s$^{-1}$. A clear spectral evolution was observed in the
outburst decay of a sample of transient X--ray pulsars, such as V0332+53 
(Makishima \etal 1990) and EXO2030+375 (Reynolds, Parmar \& White 1993).  
However in these cases the lowest detected luminosities were
substantially higher ($\approx 10^{36}$~ergs~s$^{-1}$) than in the case
of 2E0050.1--7247; this suggests that their neutron stars were further 
away from the regime of centrifugal inhibition of accretion 
which occurs close to the end of a transient X--ray pulsar 
outburst (Stella, White \& Rosner 1986).  
By requiring that the centrifugal barrier is open for an accretion
luminosity of $\sim 10^{36}$~ergs~s$^{-1}$ we derive an upper limit of 
$B< 1.5\times 10^{12}$~G to the magnetic field of the neutron star in
2E0050.1--7247.  For a spin period of 8.9~s the luminosity gap across the
centrifugal barrier should be a factor of $\sim 700$ (Corbet 1996),
indicating that when the source was a factor of $\sim 20$ fainter the
centrifugal barrier could not be completely closed and accretion onto the
neutron star still took place. The case of 2E0050.1--7247, together with
that of the other X--ray pulsars discussed above, suggests that the
production of very soft X--ray spectra results (for some yet unknown
reason) from the accreting neutron star approaching the regime of
centrifugal inhibition of accretion.  

\acknowledgments

We are grateful to A. Pizzella and L. Reduzzi for obtaining the optical
spectra of AV111.  This work was partially supported through ASI grants.

\vfill
\newpage

\figcaption{ Power spectrum of the 0.1--2.4 keV ROSAT PSPC light curve of 
1WGA J0051.8--7231 (panel $a$). The preliminary 
$3\sigma$ detection threshold is also shown. Peaks around multiples of 
$\sim 1.7 \times 10^{-4}$~Hz and $\sim 2.5\times10^{-3}$~Hz are spourious 
(see text for details). The peaks around 0.113~Hz revealed the presence of 
a 8.9~s periodic signal in 1WGA J0051.8--7231. 
The latter peaks are shown in greater detail (panel $b$) together with  
the power spectrum of the window function (panel $c$) 
Folded light curves at the best period of 
8.8816~s are shown as an insert in panel $a$ for two different energy 
ranges.} 

\figcaption{ ROSAT PSPC spectrum of 1WGA J0051.8--7231 
during the 1993 May 9--12 and 1991 Oct 8--9 observations (upper panel). 
The best fit power law models are shown, together with the corresponding 
residuals. The lower panel gives the 
1, 2 and $3\sigma$ confidence contours in the  
$N_H$ -- $\Gamma$ plane for the 1993 May 9--12 (solid lines) and 
1991 Oct 8--9 spectra (dash--dotted lines). The crosses indicate the best fit 
values. The vertical line represents the estimated galactic hydrogen column.}

\figcaption{ ROSAT PSPC image of the SMC region around 
2E0050.1--7247 / 1WGA J0051.8--7231. The ROSAT and Einstein best positions 
are given (crossed circles), together with their error circles 
($\sim$ 30$\arcsec$ and $\sim$ 40$\arcsec$, respectively). The filled circles 
indicate the optical positions of AV111 and the previously unclassified B 
star (Star1).}   

\vfill
\newpage

\begin{table}
\begin{center}
\begin{tabular}{rcrclrc}
\multicolumn{7}{c}{Table~1: ROSAT PSPC observations of 2E0050.1--7247}\\ \\
\tableline \tableline \\
\multicolumn{2}{c}{Start Time}& \multicolumn{2}{c}{Stop Time }& Seq. & 
 Exp. & Count Rate \\
\multicolumn{4}{c}{}&(\#) & 
(s) &(10$^{-3}$cts~s$^{-1}$) \\ \\[-3mm] \tableline \\[-3mm]
91 Oct 8 & 03:10 & 91 Oct 9 & 02:38 & 600195 & 16981 
&  $6.9 \pm 0.5$\\
91 Oct 10 & 03:06 & 91 Oct 10 & 04:47& 600196 & 1346 
&   $< 30$~~ $^{(a)}$ \\
92 Apr 10 & 15:30 & 92 Apr 25 & 16:41 & 600196 & 22680 
&  $< 5$~~ $^{(b)}$ \\
92 Apr 17 & 17:01 & 92 Apr 27 & 16:28 & 600195 & 9651 
& $< 10$~~ $^{(c)}$ \\     
93 May 9 & 07:17 & 93 May 12 & 20:14 & 600453 & 17599 
&  $114\pm 3$ \\
93 Nov 5 & 23:21 & 93 Nov 9 & 02:37 & 500249 & 19262 
&  $< 20$~~ $^{(d)}$ \\
93 Nov 30 & 05:39 & 93 Nov 30 & 07:37 & 500251 & 2093  
&   $< 60$~~ $^{(e)}$ \\
\tableline
\end{tabular} 
\end{center}

\tablenotetext{(a)}{ The wobble 
direction was such that the source was almost always obscured by a rib.}
\tablenotetext{(b)}{ As a result of the wobble the source was obscured by 
a rib for $\sim$50\% of the time.}
\tablenotetext{(c)}{ The wobble direction was such that the source moved 
parallel to the rib $\sim$1$\arcmin$ away from it and therefore was unobscured.}
\tablenotetext{(d)}{ As a result of the wobble the source was obscured  for 
$\sim$25\% of the time.}
\tablenotetext{(e)}{ The wobble was such that the source spent $\sim$$90$\% 
of the time under a rib or out of the field of view.}

\end{table}

\begin{table}
\begin{center}
\begin{tabular}{lcccc}
\multicolumn{5}{c}{Table~2: ROSAT PSPC and Einstein IPC Spectral Results for 
2E0050.1--7247}\\ \\
\tableline \tableline \\
& \multicolumn{2}{c}{ROSAT PSPC} & \multicolumn{2}{c}{Einstein IPC} \\
Parameter \ \ \ \ \ \ \ \ \ \ \ \ \ \ \ Seq. (\#)
 & 600453 & 600195$^a$ & 6755 & 7988 \\ \\[-3mm]
\tableline \\[-3mm]
$N_H$ (10$^{20}$ cm$^{-2}$) & 8 $\pm^3_2$ & 7 $\pm^9_3$ 
      & 8 (fixed) & 8 (fixed) \\
$\Gamma$ & 1.1 $\pm 0.2 $ & 2.9 $\pm_{0.7}^{0.8}$ 
      & 0.9 $\pm$ 0.9& 1.0 $\pm^{1.1}_{0.9} $\\
$F_x$ (10$^{-12}$ erg s$^{-1}$ cm$^{-2}$) & 2.5 & 0.1 
      & 3.4 & 0.7 \\
Count rate (10$^{-3}$ ct\,s$^{-1}$) & 114 $\pm$ 3 & 6.9 $\pm$ 0.5  
      & 70 $\pm$ 10 &16 $\pm$ 1  \\
$L_x$ (10$^{35}$ erg s$^{-1}$, d=65~kpc) & 12 & 0.5  & 14 & 3.1\\
Reduced chisquare & 1.1 & 1.0 & 0.6 & 0.5  \\[2mm]
\tableline
\end{tabular} 
\tablenotetext{a}{Oct 1991 data only.}
\end{center}

\noindent\tablecomments{X--ray fluxes and luminosities are unabsorbed 
and refer to the 0.1--2.0 keV and 0.15--3.5 keV band for the ROSAT PSPC 
and Einstein IPC, respectively. Quoted uncertainties are 
68\% confidence for one parameter of interest.}

\end{table}

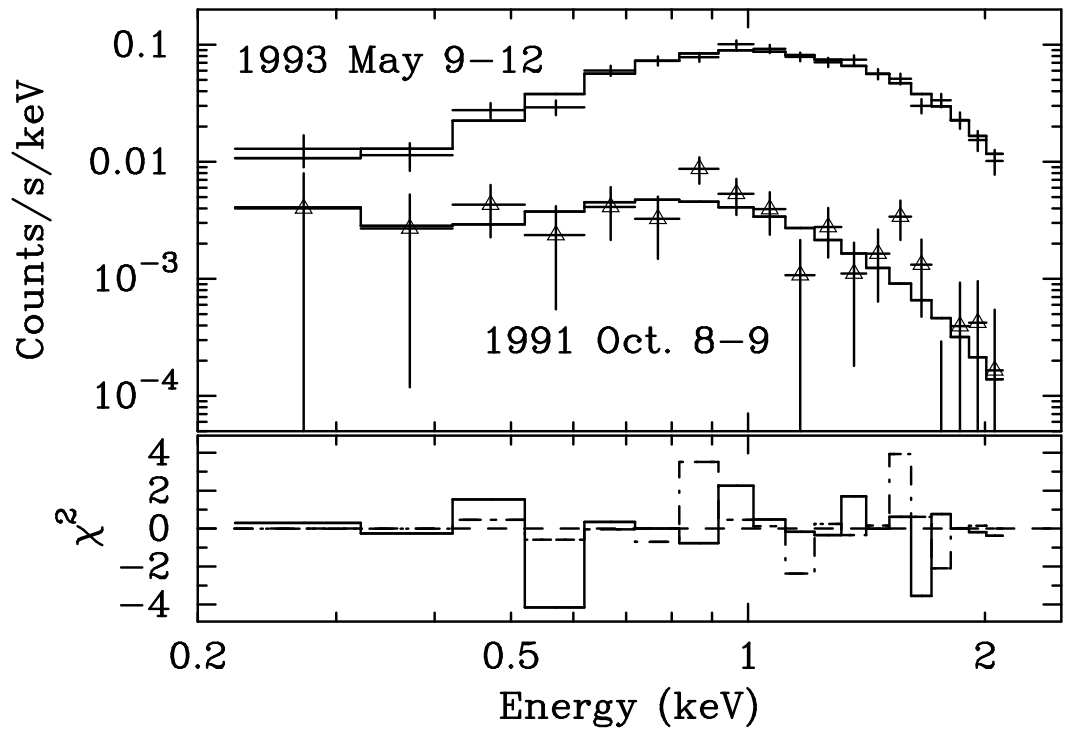
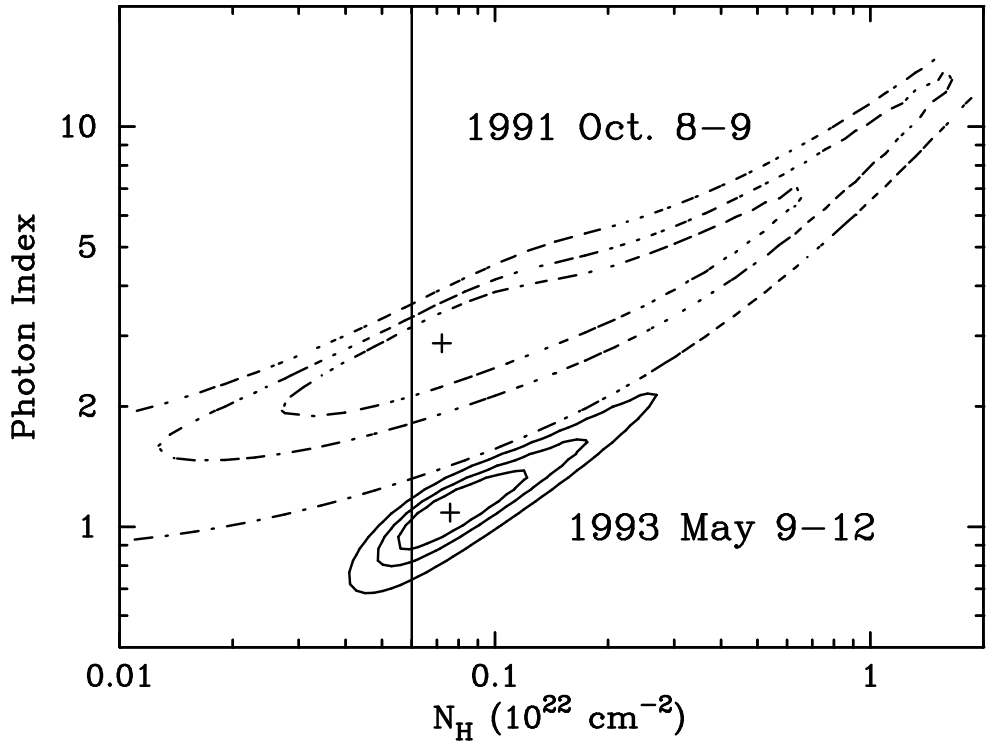
\end{document}